\documentclass{aa}

\usepackage{graphicx}
\usepackage{txfonts}
\usepackage{hyperref}
\usepackage{multirow}

\begin{document} 

   \title{Analysis of \textit{Fermi} gamma-ray burst duration distribution}

        \author{M. Tarnopolski
               \inst{1}
        }

        \institute{Astronomical Observatory, Jagiellonian University,
                Orla 171, 30-244 Cracow\\
                \email{mariusz.tarnopolski@uj.edu.pl}
        }

        \date{Received; accepted}

\abstract
{Two classes of gamma-ray bursts (GRBs), short and long, have been determined without any doubts, and are usually prescribed to different physical scenarios. A third class, intermediate in $T_{90}$ durations has been reported  in the datasets of BATSE, {\it Swift}, RHESSI, and possibly BeppoSAX. The latest release of $>1500$ GRBs observed by {\it Fermi} gives an opportunity to further investigate the duration distribution.}
{The aim of this paper is to investigate whether a third class is present in the $\log T_{90}$ distribution, or whether it is described by a bimodal distribution.}
{A standard $\chi^2$ fitting of a mixture of Gaussians was applied to 25 histograms with different binnings.}
{Different binnings give various values of the fitting parameters, as well as the shape of the fitted curve. Among five statistically significant fits, none is trimodal.}
{Locations of the Gaussian components are in agreement with previous works. However, a trimodal distribution, understood in the sense of having three distinct peaks, is not found for any binning. It is concluded that the duration distribution in the {\it Fermi} data is well described by a mixture of three log-normal distributions, but it is intrinsically bimodal, hence no third class is present in the $T_{90}$ data of {\it Fermi}. It is suggested that the log-normal fit may not be an adequate model.}


   \maketitle

\section{Introduction}

Gamma-ray bursts (GRBs) are most commonly classified based on their duration time $T_{90}$ (time during which 90\% of the burst fluence is accumulated, starting from the time at which 5\% of the total fluence was detected). \citet{mazets} first observed the bimodal distribution of $T_{90}$ drawn for 143 events detected in the KONUS experiment. \citet{kouve} also found a bimodal structure in the $\log T_{90}$ distribution in BATSE 1B dataset; based on this data GRBs are divided into short ($T_{90}<2\,{\rm s}$) and long ($T_{90}>2\,{\rm s}$) classes. \citet{horvath98} and \citet{mukh} independently discovered a third peak in the duration distribution in the BATSE 3B catalog \citep{meegan96}, located between the short and long peaks, and the statistical existence of this intermediate class was supported \citep{horvath02} with the use of BATSE 4B data. Evidence for a third component in $\log T_{90}$ was also found in the {\it Swift} data
\citep{horvath08a,horvath08b,zhang,huja,horvath10}. Other datasets, i.e., RHESSI \citep{ripa} and BeppoSAX \citep{horvath09}, are both in agreement with earlier results regarding the bimodal distribution, and the detection of a third component was established on a lower significance level (compared to BATSE and {\it Swift}) owing to less populated samples. Hence, four different satellites provided hints about the existence of a third class of GRBs. Contrary to this, the durations  observed by INTEGRAL have a unimodal distribution that extends to the shortest timescales as a power law \citep{savchenko}.

Only one dataset (BATSE 3B) was truly trimodal in the sense of having three peaks. In the others, a three-Gaussian distribution was found to follow the histogram better than a two-Gaussian, but those fits yielded only two peaks, so although statistical analyses support the presence of a third class, its existence is still elusive and may be due to selection and instrument effects, or might be described by a distribution that is not necessarily a mixture of Gaussians. The latest release is due to {\it Fermi} GBM observations \citep{gruber,kienlin} and consists of 1568 GRBs\footnote{\url{http://heasarc.gsfc.nasa.gov/W3Browse/fermi/fermigbrst.html}, accessed on March 12, 2015.} with calculated $T_{90}$. This sample, to the best of the author's knowledge, has not yet been investigated for the presence of an intermediate class; only \citet{horvath12} and \citet{qin} conducted research on a subsample consisting of 425 bursts from the first release of the catalog.

Previous analyses, although consistent with each other (except INTEGRAL), showed variances in locations of the Gaussian components, their dispersion, relative heights, and statistical significance of a trimodal fit compared to a bimodal distribution. The aim of this article is to perform a statistical analysis of the {\it Fermi} database in order to verify the existence of the intermediate-duration GRB class. The current sample constitutes $\sim 83\%$ of long GRBs ($T_{90}>2\,{\rm s}$), which is a larger percentage than in previous catalogs. A conventional $\chi^2$ fitting is applied. Since BATSE 4B \citep{horvath02}, the maximum log-likelihood method has been used. In the case of the {\it Swift} data this was fully justified because of the small size of the population \citep{horvath08a,horvath08b}. However, in the first analysis of the BATSE 3B data \citep{horvath98}, which consisted of 797 GRBs, three classes were detected with the $\chi^2$ method, which was later supported by the maximum log-likelihood method on a bigger dataset \citep{horvath02}. Hence, the number of GRBs under consideration in this study should not be too small to conduct a conventional fitting analysis. To verify whether the $\chi^2$ fitting is applicable, the BATSE 4B data (2041 GRBs) were re-examined and results similar to those of \citet{horvath02} were obtained. To be sure that the method chosen does not rule out any feature in the data under consideration, the maximum log-likelihood was also applied, but since the results were not different from those obtained using the $\chi^2$ fitting, they are not reported here  or mentioned  in what follows.

This article is organized as follows. Section 2 presents the $\chi^2$ fittings of different models. In Sect. 3 a comparison with previous results is discussed. Section 4 is devoted to discussion, and Sect. 5 gives concluding remarks. The computer algebra system \textsc{Mathematica}\textsuperscript{\textregistered} v10.0.2 is applied throughout this paper.

\section{The $\chi^2$ fittings}

\subsection{Sample}

In this section a dataset of 1566 GRBs is used. Two durations (the shortest and the longest) were excluded because, for every binning, they were separated from the rest of the distribution by empty bins.

\subsection{Standard normal distributions}

A least squares fitting of a multi-component log-normal distribution to the dataset of durations $T_{90}$ is performed, i.e., a mixture of Gaussians
\begin{equation}
f_k=\sum\limits_{i=1}^k A_i \mathcal{N}_i(\mu_i,\sigma^2_i)=\sum\limits_{i=1}^k \frac{A_i}{\sqrt{2\pi}\sigma_i}\exp\left(-\frac{(x-\mu_i)^2}{2\sigma^2_i}\right)
\label{eq1}
\end{equation}
is fitted to a histogram of $\log T_{90}$. A significance level of \mbox{$\alpha=0.05$} is adopted. Twenty-five binnings are applied, defined by the bin widths $w$ from 0.30 to 0.06 with a step of 0.01. The corresponding number of bins ranges from 15 to 69. A number of binnings is chosen rather than a neutral bin width established according to some conventional rule (e.g., Freedman-Diaconis, Scott, Knuth) because of \citet{koen}, who found two fits to {\it Swift} data that could not be rejected, hence restricting the analysis to only one binning might be concealing. For each binning the same fitting procedure is performed as described.

The $\chi^2$ of a fit is calculated in a standard way as
\begin{equation}
\chi^2=\sum\limits_{i=1}^N\frac{(O_i-E_i)^2}{E_i},
\label{eq2}
\end{equation}
where $O_i$ is the observed $i$-th value, $E_i$ is the value expected based on the fit, and $N$ is the number of bins. The number of degrees of freedom, $\mbox{\textit{dof}}$, of the $\chi^2$ test statistic is $\mbox{\textit{dof}}=N-m-1$, where $m$ is the number of parameters used; for a $k$-Gaussian, $m=3k$. First, a single-Gaussian fit is performed for all binnings. The $\chi^2$ values range from $12\,180$ to $54\,548$, with $p$-values being numerically equal to zero in all cases. It follows from the huge $\chi^2$ values that it is extremely unlikely that the single-Gaussian  fit describes the $\log T_{90}$ distribution correctly.

\begin{figure}
\centering
\includegraphics[width=\hsize]{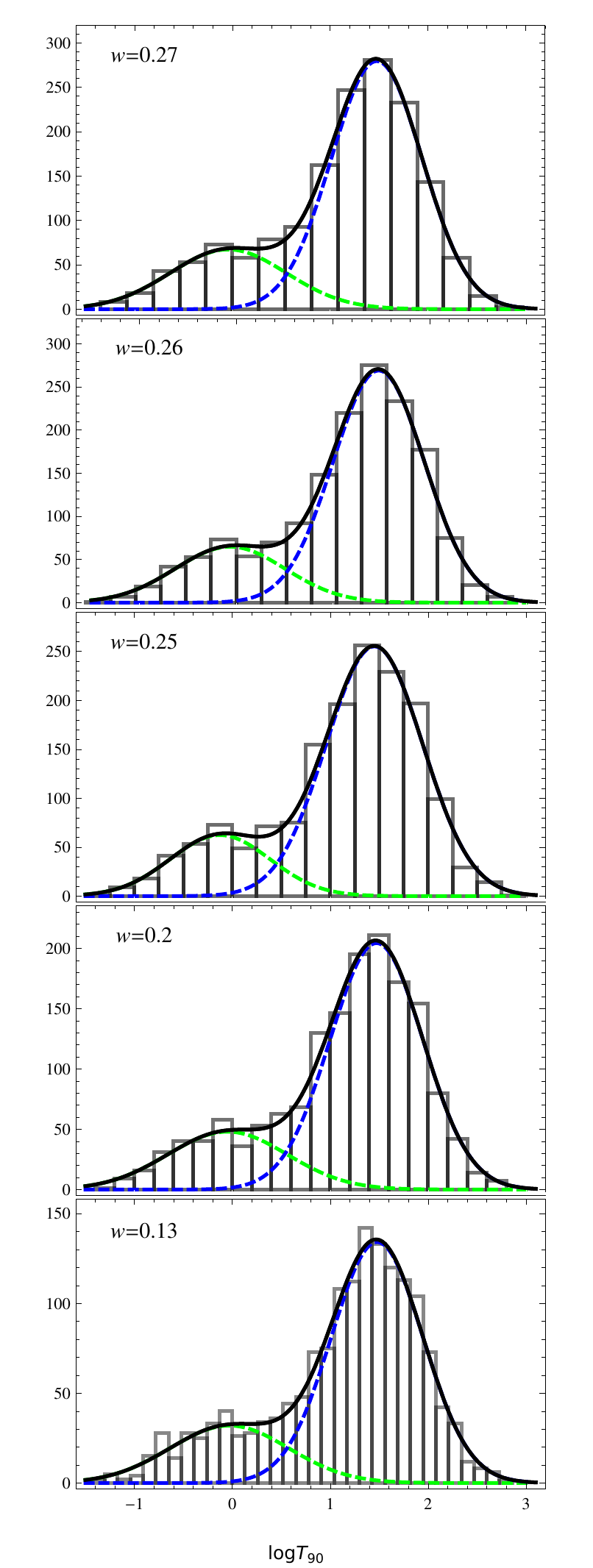}
\caption{Two-Gaussian fits (solid black)  to $\log T_{90}$ distributions. All except $w=0.25$ are statistically significant. Colored dashed curves are the components of the mixture distribution.}
\label{fig1}
\end{figure}

Next, a two-Gaussian fit is performed and the resulting $\chi^2$ are much lower; however, the majority of $p$-values indicates that the distributions do not follow the data well at the $\alpha=0.05$ significance level. Fitted parameters are gathered in Table~\ref{tbl1}, with $p$-values greater than $\alpha$ written in bold, while the curves are displayed in Fig.~\ref{fig1}. The bimodal structure is well represented by a two-Gaussian fit in four out of five cases displayed in Fig.~\ref{fig1}. The statistically unsignificant fit for $w=0.25$ is also shown due to a comparison with a three-Gaussian fit to be performed further on.
\begin{table}
\caption{Parameters of a two-Gaussian fit}
\label{tbl1}
\centering
\begin{tabular}{c c c c c c c}
\hline\hline
  $w$ & $i$ & $\mu_i$ & $\sigma_i$ & $A_i$ & $\chi^2$ & $p$-val \\
  \hline
\multirow{2}{*}{0.27} & 1 & -0.042 & 0.595 & 100.2 & \multirow{2}{*}{6.467} & \multirow{2}{*}{\bf 0.692} \\
     & 2 & 1.477 & 0.465 & 325.7 & & \\
  \hline
\multirow{2}{*}{0.26} & 1 & -0.063 & 0.569 & 92.30 & \multirow{2}{*}{12.23} & \multirow{2}{*}{\bf 0.270} \\
     & 2 & 1.475  & 0.473 & 318.8 & & \\
  \hline
\multirow{2}{*}{0.25} & 1 & -0.125 & 0.510 & 79.55 & \multirow{2}{*}{22.00} & \multirow{2}{*}{0.024} \\
     & 2 & 1.453  & 0.494 & 316.1 & & \\
  \hline
\multirow{2}{*}{0.20} & 1 & -0.049 & 0.611 & 73.34 & \multirow{2}{*}{22.09} & \multirow{2}{*}{\bf 0.077} \\
     & 2 & 1.473  & 0.476 & 243.6 & & \\
  \hline
\multirow{2}{*}{0.13} & 1 & -0.030 & 0.607 & 48.56 & \multirow{2}{*}{33.74} & \multirow{2}{*}{\bf 0.142} \\
     & 2 & 1.480  & 0.468 & 157.2 & & \\
  \hline
\end{tabular}
\end{table}

\begin{figure}
\centering
\includegraphics[width=\hsize]{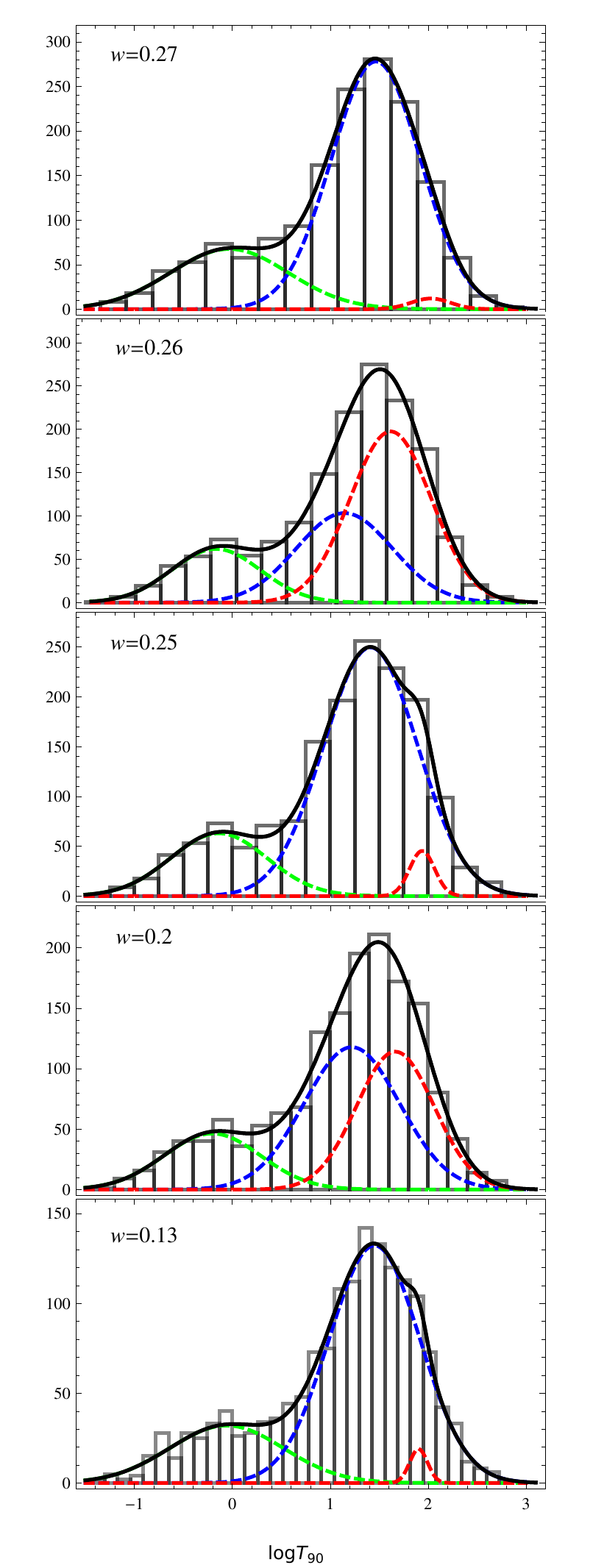}
\caption{The same as Fig~\ref{fig1}, but for a three-Gaussian fit.}
\label{fig2}
\end{figure}

Finally, a three-Gaussian fit is performed in the same manner. The resulting $\chi^2$ are naturally lower than they were in the previous case. Again, a majority of the corresponding $p$-values that are smaller than $\alpha$ indicates that in general the duration distribution is not well described by a mixture of three log-normal components. Parameters of the five fits with $p\textrm{-values} >\alpha$ are gathered in Table~\ref{tbl2}, and the fitted curves are displayed in Fig.~\ref{fig2}. It is important to note that these five fits are all statistically significant, including the fit for $w=0.25$, for which a hypothesis that the histogram is well described by a two-Gaussian fit was rejected.
\begin{table}
\caption{Parameters of a three-Gaussian fit}
\label{tbl2}
\centering
\begin{tabular}{c c c c c c c}
\hline\hline
  $w$ & $i$ & $\mu_i$ & $\sigma_i$ & $A_i$ & $\chi^2$ & $p$-val \\
  \hline
     & 1 & -0.030 & 0.603 & 102.0 & & \\
0.27 & 2 & 1.466  & 0.455 & 317.1 & 5.333 & {\bf 0.502} \\
     & 3 & 2.027  & 0.201 & 6.014 & & \\
\hline
     & 1 & -0.210 & 0.461 & 71.48 & & \\
0.26 & 2 & 1.119  & 0.450 & 128.6 & 6.819 & {\bf 0.448} \\
     & 3 & 1.598  & 0.421 & 208.4 & & \\
\hline
     & 1 & -0.137 & 0.492 & 77.73 & & \\
0.25 & 2 & 1.414  & 0.480 & 300.4 & 13.52 & {\bf 0.095} \\
     & 3 & 1.939  & 0.128 & 14.49 & & \\
\hline
     & 1 & -0.204 & 0.493 & 57.30 & & \\
0.20 & 2 & 1.221  & 0.488 & 144.0 & 17.71 & {\bf 0.087} \\
     & 3 & 1.665  & 0.396 & 113.1 & & \\
\hline
     & 1 & -0.058 & 0.581 & 46.61 & & \\
0.13 & 2 & 1.453  & 0.464 & 153.7 & 29.42 & {\bf 0.167} \\
     & 3 & 1.903  & 0.092 & 4.328 & & \\
\hline
\end{tabular}
\end{table}

The probability that the third component originates from statistical fluctuations may be estimated by comparing a three-Gaussian  with a two-Gaussian fit, based on an observation \citep{band} that 
\begin{equation}
\Delta\chi^2=\chi^2_1-\chi^2_2\stackrel{\textbf{\textrm{.}}}{=}\chi^2(\Delta\nu),
\label{eq3}
\end{equation}
where $\Delta\nu=\nu_2-\nu_1$ is the difference in the degrees of freedom of fits under consideration, equal to three when $\Delta k=1$ (compare with Eq.~(\ref{eq1})), and $\stackrel{\textbf{\textrm{.}}}{=}$ denotes equality in distribution.

To elaborate to what degree a three-Gaussian fit is better than a two-Gaussian, Eq.~(\ref{eq3}) is applied and the $p$-values are inferred from a $\chi^2$ distribution with three degrees of freedom. A small \mbox{$p$-value} indicates a small probability that the two-Gaussian model alone describes the data in comparison with a three-Gaussian. The results, gathered in Table~\ref{tbl3}, indicate that in all binnings there is a $>20\%$ probability (exceeding $50\%$ for the remaining three statistically significant cases) that a three-Gaussian is a significant improvent over a two-Gaussian fit. This does not mean that the three-Gaussian models are a good description of the data, as the $\chi^2$ from Table~\ref{tbl2} lead to rejection of the null hypothesis for only one out of five binnings with $p$-values greater than $\alpha$ in the three-Gaussian case. For this unsignificant fit, the $p$-value computed from $\Delta\chi^2$ means that a bad three-Gaussian is better than a bad two-Gaussian. For the fits that are statistically significant (when fitting both a two- and a three-Gaussian) the conslusion is that  a  null hypothesis cannot be rejected for both, but a three-Gaussian describes the data better with a probability $>50\%$. Nevertheless, this probability is insufficient to claim a detection;  in \citep{horvath09} it was concluded that even a $96\%$ significance level is too small to be considered  evidence.
\begin{table}
\caption{Improvements of a three-Gaussian over a two-Gaussian fit}
\label{tbl3}
\centering
\begin{tabular}{c c c}
\hline\hline
$w$ & $\Delta\chi^2$ & $p$-value \\
\hline
0.27 & 1.134 & {\bf 0.767} \\
0.26 & 5.411 & {\bf 0.144} \\
0.15 & 8.480 & 0.037 \\
0.20 & 4.380 & {\bf 0.223} \\
0.13 & 4.320 & {\bf 0.229} \\
\hline
\end{tabular}
\end{table}

The {\it Fermi} distribution is dominated by long GRBs, which constitute $\sim 83\%$ of the sample, that is manifested through a significantly higher dispersion of the short GRB distribution. Moreover, it was proposed that the distribution for short GRBs was nearly flat for $T_{90}\lesssim 2\,{\rm s}$ \citep{bromberg2,bromberg} (compare with \citep{savchenko}); however, for smaller bin widths a statistical noise starts to dominate. This is supported by a very small $p$-value of \mbox{$\sim 10^{-4}-10^{-7}$} for the smaller bin widths, $w=0.11-0.06$, which indicates that at these binnings statistical noise is dominating.

Three out of five statistically significant fittings (\mbox{$w=0.27,\,0.25,\,0.13$}) located the third component beyond the main peak for long GRBs. One fit ($w=0.26$) showed an excess at $T_{90}\approx 10\,{\rm s}$ that might be assigned to an intermediate class; however, the fitted curve is very similar to a two-Gaussian. The last fit, with $w=0.20$, detected two components (in addition to a peak related to short GRBs) of approximately the same height and comparable standard deviations in a region of long GRBs. For these  last two fits, the dispersion of a corresponding peak of a two-Gaussian is greater than the dispersion of both the components in the three-Gaussian, hence it gives a hint toward the bimodality of the GRBs \citep{schill}. This is in agreement with a very high probability that the third group in a three-component model is a statistical fluctuation.

\section{Comparison with former results}

A detailed comparison between BATSE 4B and {\it Swift} catalogs has been conducted \citep{huja} and the results were found to be consistent, also in the means of differences between the instruments. The RHESSI observations were also taken into account \citep{ripa} and roughly the same distribution as in previous catalogs was reported. The BeppoSAX data were of a relatively low population and the analysis showed the presence of three components on a significance level lower than in previous catalogs, but two classes -- the intermediate and long -- were detected with high certainty \citep{horvath09}. As short GRBs were underpopulated, the three-Gaussian fit, despite following the observations better than a two-Gaussian, consisted of a component with a high dispersion. Overall, classification analyses found three components in all four samples. Interestingly, the dataset from INTEGRAL \citep{savchenko} yields a unimodal distribution. The latest well-populated sample, based on the {\it Fermi} data, was investigated some time ago \citep{horvath12} using PCA and multiclustering analysis, and a three-group structure was found in a multidimensional parameter space including duration, total fluence, hardness ratio, and peakflux256. Hence, data from five satellites supported a three-component distribution of GRBs by means of statistical significance.

\begin{figure}
\centering
\includegraphics[width=\hsize]{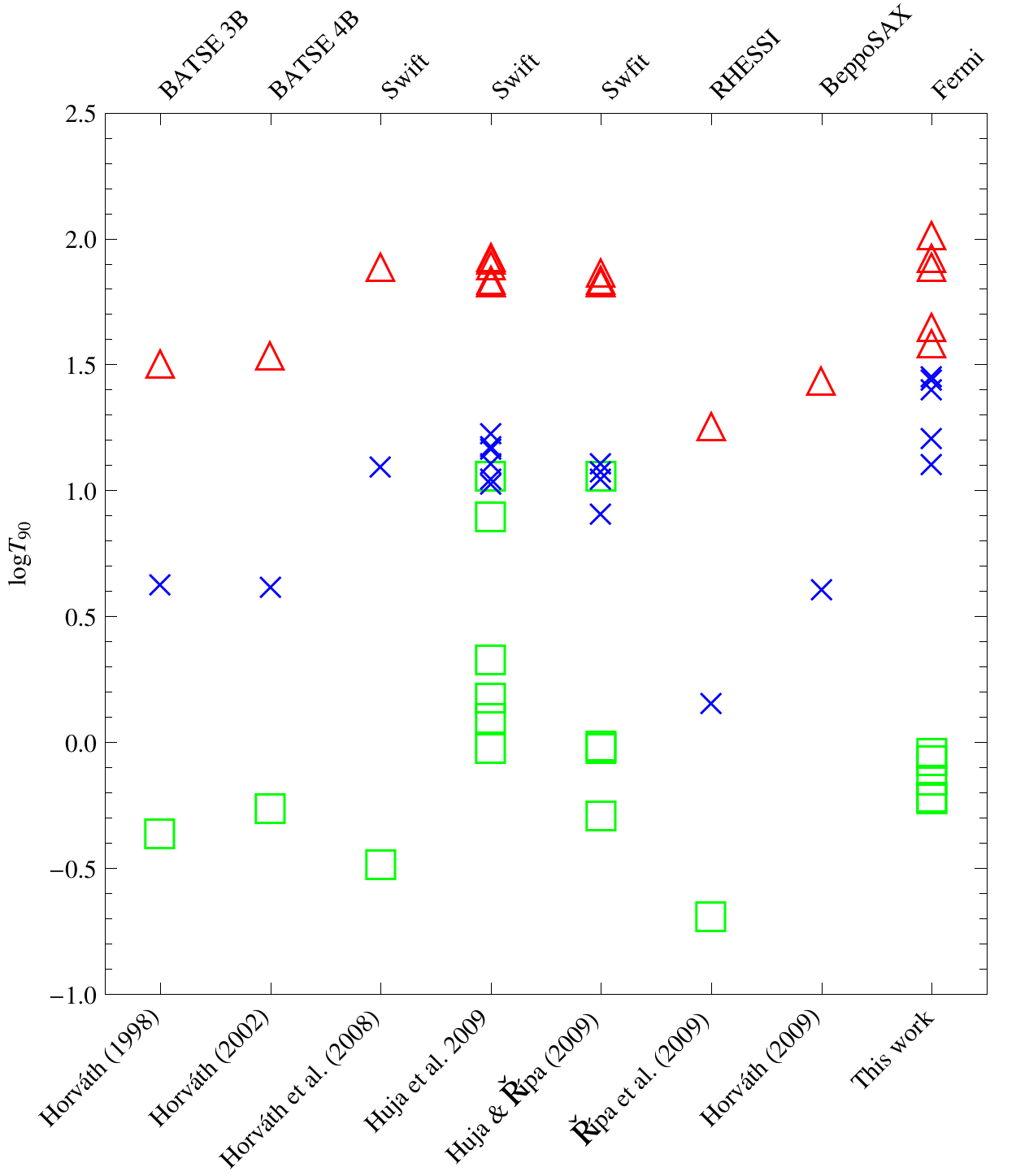}
\caption{Locations of the short (green squares), intermediate (blue crosses) and long (red triangles) GRBs from previous research and this work.}
\label{fig3}
\end{figure}

In Fig.~\ref{fig3} all of the components' locations found by the above-mentioned univariate analyses are plotted. Results from this work, i.e., locations of the three-Gaussian components found by a standard log-normal fitting, are consistent with previous results by means of locations as well as relative separations -- mean durations of the five statistically significant fits are centered at $0.745\,{\rm s}$, $21.61\,{\rm s}$, and $67.05\,{\rm s}$ for short, intermediate, and long GRBs, respectively. The current sample of 1566 GRBs comprises 17\% of short GRBs and 83\% of long ones, based on the conventional classification $\lessgtr 2\,{\rm s}$. Because of an insufficient separation of the intermediate and long components, it is impossible to conclude what  the population of intermediate class GRBs in {\it Fermi} data is, as the distribution is apparently bimodal and shows no evidence for the third class being present, and one can associate the  {\it intermediate} and {\it long} groups with a single class. It is important to note that in {\it Fermi} the sensitivity at very soft and very hard GRBs was higher than in BATSE \citep{meegan09}. Soft GRBs are intermediate in duration, and hard GRBs have short durations. Hence, an increase in intermediate GRBs relative to long ones might be expected as a consequence of improving instruments, yet the third class remains elusive. {\it Swift} is more sensitive in soft bands than BATSE was, hence its dataset has a low fraction of short GRBs. Therefore, the group populations inferred from {\it Fermi} observations are reasonable considering the characteristics of the instruments.

\section{Discussion}

Among the 25 three-Gaussian fittings performed, five (\mbox{$w=0.27,\,0.26,\,0.25,\,0.20,\,0.13$}) turned out to be statistically significant, with $p$-values exceeding the significance level \mbox{$\alpha=0.05$}. Locations of the respective groups for the five fits are close to each other with means $\mu_1=-0.128\pm 0.082$ (short GRBs), $\mu_2=1.335\pm 0.156$ and $\mu_3=1.826\pm 0.185$ (long GRBs), the error being the standard deviation of the average (compare with Fig.~\ref{fig3}). Previous works on datasets from BATSE \citep{horvath98,horvath02} and {\it Swift} \citep{horvath08a,huja} indicated that a three-Gaussian is a better fit than a corresponding two-Gaussian. On the other hand, a three-Gaussian fit to RHESSI \citep{ripa} data yielded only a $93\%$ probability of being correct compared to a two-Gaussian, meaning that there is a remarkable $7\%$ probability that the $\log T_{90}$ is well described by a two-Gaussian, while for BeppoSAX \citep{horvath09} the goodness-of-fit was not reported (only the maximum log-likelihoods).

Moreover, the two greater means of the three-Gaussian from Fig.~\ref{fig2} ($w=0.20$) do not satisfy the criterion of bimodality \citep{schill}
\begin{equation}
|\mu_A-\mu_B|>(\sigma_A+\sigma_B)\cdot S\left(\frac{\sigma_A^2}{\sigma_B^2}\right),
\label{eq4}
\end{equation}
with $\sigma_A\leq\sigma_B$, and $S(r)$ being the separation factor equal to 0.98 in the case of $w=0.20$ (assuming equal weights; arbitrary mixtures yield an even higher value of the factor) for $\sigma_2$ and $\sigma_3$ from Table~\ref{tbl2}, hence long GRBs ($T_{90}>2\,{\rm s}$) do not have a bimodal distribution and are described by a single peak, meaning that the distribution over the whole range of available durations $T_{90}$ is bimodal, with peaks corresponding to short and long GRB populations. The fit for $w=0.26$  does not fulfil this condition either, where the appropriate $S(r)=1.16$ is taken from \citep{schill}. The remaining three cases, although the shoulder is prominent, are also apparently bimodal.

The relative improvements (Table~\ref{tbl3}) indicate an enourmous probability, ranging from $14\%$ to $77\%$, that the third component in a three-Gaussian fit is a chance occurrence compared to a two-Gaussian.

Finally, among previous research, only the BATSE 3B data revealed a truly trimodal $\log T_{90}$ distribution, i.e., having three local maxima, which was not present in the following release, BATSE~4B (where only a bump was present), and is non-existent in the current BATSE catalog.  {\it Swift} also observed two local maxima, although with a prominent shoulder on the left side of the long GRB peak \citep{horvath08a}, detected by means of the maximum log-likelihood method. \citet{huja} also obtained a bimodal distribution with a bump on one side of the long GRB peak, although somewhat weaker. The explanation may be that for the sample of 388 GRBs, a maximum log-likelihood method is more robust than applying a $\chi^2$ fitting. The latter may be a drawback in undersampled populations. Fortunately, this is not a case in the {\it Fermi} data. The RHESSI distribution \citep{ripa} is also characterized by a bimodal fit with a shoulder, while in BeppoSAX  clearly separated intermediate and long groups were found \citep{horvath09}; however, no short GRBs were detected in the $\log T_{90}$ distribution there. This is most likely due to a low trigger efficiency to short GRBs, which are highly underpopulated in that sample. Interestingly, an early analysis of 222 GRBs from {\it Swift} \citep{horvath08b} detected the short class, while the long GRBs were unimodal, again with a bump. Finally, the standard three-Gaussian fits passed the Anderson-Darling test, hence a mixture of log-normal distributions account well for the observed durations.

The distributions fitted in this paper are strongly dependent on the binning applied when the locations of components and their relative amplitudes are considered. No statistically significant trimodal fit was found, although a shoulder (in three out of five fits) was detected beyond the region where previous works found the intermediate class, which is a surprising result. Still, a three-Gaussian fit is a better fit than a two-Gaussian, statistically speaking. However, it is arguable whether this confirms the existence of the third class of GRBs. As the sum of two normal distributions is skewed, which is the apparent case in the {\it Fermi} sample, the underlying multimodal distribution may not necessarily be composed of Gaussians. Recently, \citet{zitouni} suggested that the duration distribution corresponding to the collapsar scenario (associated with long GRBs) might not  necessarily be symmetrical  because of  a non-symmetrical distribution of envelope masses of the progenitors.

Moreover, the more components the mixture distribution has, the more parameters are available for the curve to fit the histogram. This may be the primary reason for the calculated goodness-of-fit. On the other hand, in terms of statistical accuracy, the hypothesis that three fundamental components are needed to describe observed durations is corroborated. The true underlying form of the distribution remains obscure. Short GRBs have been underpopulated in observations ever since, and some instrument biases have been proposed to account for the emergence of a third peak. The trimodality in the BATSE 3B catalog was greatly diminished in the 4B version, which may be directly attributed  to collecting a more complete sample. The {\it Fermi} database contains 75\% of the number of GRBs observed in BATSE current catalog, hence further observations, as well as theoretical models that could account for the diversity of GRB events in more detail may clarify the properties of the $T_{90}$ distribution and determine whether an intermediate class of GRBs is a physical or a statistical phenomenom.

Finally, a mixture of log-normal Gaussians may not be a proper model for the $T_{90}$ distribution, and a mixture of intrinsically skewed distributions may be a better explanation of the observed features of the histogram. On the other hand, the duration distribution might not be a sum of components defined on $(-\infty,+\infty)$, but might be a piecewise function.

\section{Conclusions}

\begin{enumerate}
\item A mixture of three standard normal distributions was found to be statistically significant in describing the $\log T_{90}$ distribution for 1566 {\it Fermi} GRBs for bin widths \mbox{$w=0.27,\,0.26,\,0.25,\,0.20,\,0.13$}. Average locations of the components are equal to $0.745\,{\rm s}$, $21.61\,{\rm s}$, and $67.05\,{\rm s}$ for short, intermediate, and long GRBs, respectively, or $-0.128$, $1.335$, and $1.826$ in log-scale. These results are in agreement with values obtained from previous catalogs: BATSE, {\it Swift}, RHESSI, and BeppoSAX.
\item The relative improvements of a three-Gaussian fit over a two-Gaussian imply that there is a significant probability, varying from $14\%$ to $77\%$ among the fits, that the third component is a chance occurrence. Therefore, the third GRB class is unlikely to be present in the {\it Fermi} data.
\item None of the fits is trimodal (in the sense of having three distinct peaks). Although three out of five fits show a prominent shoulder on the {\it \emph{right}} side of the long GRB peak, the evidence is not sufficient to claim detection of a third class in the {\it Fermi} data. Therefore, the distribution of {\it Fermi} durations is intrinsically bimodal, hence no evidence for an intermediate class of GRBs has been found.
\item The observed asymmetry may come from an underlying distribution composed of two skewed components, or it may be a piecewise function.
\end{enumerate}

\begin{acknowledgements}
The author acknowledges fruitful discussions with Micha\l\ Wyr\c ebowski and Arkadiusz Kosior, and wishes to thank the anonymous referee for useful comments that led to significant improvements of the paper.
\end{acknowledgements}

\end{document}